\documentclass[preprintnumbers, prd, floatfix,twocolumn,preprintnumbers, letterpaper, superscriptaddress,nofootinbib]{revtex4-1}
\usepackage{amsfonts}
\usepackage{amsmath}
\usepackage{amssymb,epsf}
\usepackage{latexsym}
\usepackage{graphicx,epsfig}
\usepackage{amssymb}
\usepackage{float}
\usepackage{subfigure}
\usepackage[colorlinks=true,citecolor=blue,linkcolor=blue,urlcolor=black]{hyperref}
\usepackage{dcolumn}
\usepackage{psfrag}
\usepackage{wrapfig}
\usepackage{makeidx}
\usepackage{epsf}
\usepackage{color}
\usepackage{multirow}
\usepackage{tabularx}
\usepackage{array}
\newcolumntype{c}[1]{>{\centering\arraybackslash}p{#1}}

\begin{document}
\title{Barrow Cosmology and Big-Bang Nucleosynthesis}
\author{Ahmad Sheykhi \footnote{asheykhi@shirazu.ac.ir} and Ava Shahbazi}
\affiliation{Department of Physics, College of
Science, Shiraz University, Shiraz 71454, Iran \\
Biruni Observatory, College of Science, Shiraz University, Shiraz
71454, Iran}
\begin{abstract}
Using thermodynamics-gravity conjecture, we present the formal
derivation of the modified Friedmann equations inspired by the
Barrow entropy, $S\sim A ^{1+\delta/2}$, where $0\leq\delta\leq 1$
is the Barrow exponent and $A$ is the horizon area. We then
constrain the exponent $\delta$ by using Big-Bang Nucleosynthesis
(BBN) observational data. In order to impose the upper bound on
the Barrow exponent $\delta$, we set the observational bound on
$\left| \frac{\delta T_f} {T_f }\right|$. We find out that the
Barrow parameter $\delta$ should be around $ \delta \simeq 0.01$
in order not to spoil the BBN era. Next we derive the bound on the
Barrow exponent $\delta$ in a different approach in which we
analyze the effects of Barrow cosmology on the primordial
abundances of light elements i.e. Helium $_{}^{4}\textit{He}$,
Deuterium $D$ and Lithium $_{}^{7}\textit{Li}$. We observe that
the deviation from standard Bekenstein-Hawking expression is small
as expected. Additionally we present the relation between cosmic
time $t$ and temperature $T$ in the context of modified Barrow
cosmology. We confirm that the temperature of the early universe
increases as the Barrow exponent $\delta$ (fractal structure of
the horizon) increases, too.
\end{abstract}
\maketitle
\section{Introduction\label{Intro}}
After the discovery of the black hole thermodynamics as well as
four laws of black hole mechanics
\cite{Bardeen,Bekenstein,Hawking}, the profound connection between
gravity and thermodynamics has been speculated. In particular, the
derivation of the Einstein field equation of gravity from the
first law of thermodynamics by Jacobson \cite {Jacobson}, reveals
that gravity is nothing but the thermodynamics of spacetime at
large scales. The deep connection between gravity and
thermodynamics has been confirmed in different setups
\cite{Verlinde,Padman,Padmann}. In the background of
Friedmann-Robertson-Walker (FRW) universe, it has been shown that
Friedmann equations describing the evolution of the universe can
be translated to the first law of thermodynamics on the apparent
horizon \cite {Eling,Akbar, MAkbar,Cai,Asheykhi,Sheykh}. The
studies were also generalized to the modified cosmological models,
based on the modification of the Bekenstein-Hawking area law of
entropy associated with the horizon
\cite{Paranj,Jamil,Caiii,Wang,Setare,Fan,Tsheykhi}.

An interesting modification to the area law was proposed by J. D.
Barrow, who argued that black hole horizon may have a fractal structure
due to quantum fluctuations \cite {JohnD}. In this viewpoint,
quantum-gravitational effects may deform the black hole horizon
which can be described by a fractal structure as a first
approximation \cite {JohnD}. Consequently the black hole entropy
gets a deviation from the standard Bekenstein-Hawking area law, and
is modified as \cite {John}
\begin{equation} \label {S}
 S_h=\left ( \frac{A}{A_0}\right )^{1+\delta/2},
\end{equation}
where $A$ is the horizon area and $ A_{0}$ is the Planck area. The
exponent $\delta$ ranges in the interval $0\leq \delta\leq 1$, and
it shows the amount of quantum-gravitational deformation effects
\cite {John,HawkingS,Hooft}. The extreme values $\delta=1$ stands
for the most deformed and intricate structure, while for
$\delta=0$, the area law of entropy will be restored
($A_0\rightarrow 4G$) \cite{Shey1}. In the framework of
thermodynamics-gravity conjecture, it was argued that one can
derive the friedmann equations of FRW universe by applying the
first law of thermodynamics on the apparent horizon. In this
derivation, the entropy expression plays a crucial role. Any
modification to the area law, shall modify the corresponding
Friedmann equations. Therefore, by applying the first law of
thermodynamics on the apparent horizon, and using the modified
Barrow entropy \eqref{S} instead of Bekenstin-Hawking one, one can
get the modified Friedmann equations \cite{Shey1} which leads to a
so-called modified Barrow cosmology \cite{Shey2}. The Barrow
exponent $\delta$ con be constrained by using observational data.
For example, using BBN data one can impose constraints on the
exponent of Barrow entropy \cite{John}. It was shown that the
Barrow exponent should be $\delta< 1.4 \times 10^{-4}$ in order
not to spoil the BBN epoch. This implies that the deviation from
standard area law of entropy should be small as expected
\cite{John}.

BBN refers to the process that took place in the early universe,
shortly after the Big Bang, when the universe was extremely hot
and dense. BBN occurred roughly between the very first
centiseconds after the Big-Bang ($\sim 0.01$ sec) and a few
hundred seconds ($\sim 3$ minutes) after it. During BBN, the
temperature of the universe allows for nuclear reactions to occur,
leading to the formation of light elements such as $H$, $He$ and
$Li$. As the universe expanded and cooled, these reactions slowed
down and eventually stopped, resulting in the abundance of these
light elements observed today. Hence BBN, together with cosmic
microwave background radiation provide a strong evidence of the
hot, dense state of the early universe. In general, the principles
of BBN allows to impose strict constraints on various cosmological
models. The main idea is to demand consistency between theoretical
predictions, and BBN observational data. In this regard, the
deviation of the freeze-out temperature in comparison to standard
cosmological model can be calculated. This allows to compare the
theoretical results of the modified cosmological models with the
observational constraints on $\left| \frac{\delta T_f}{T_f}\right|
$\cite{John,Luciano,Anish}. A modified cosmological model can also
affect the primordial abundances of aforementioned light elements
in the early universe \cite{Anish}. It is important to note that
although the standard BBN theory successfully explains the
observed amounts of elements like Helium and Hydrogen, it
encounters significant issues when it comes to Lithium
\cite{Boran}. This leads to what is known as the
\textit{cosmological Lithium problem}, which is a highly debated
topic in the modern cosmology. It is interesting to check whether
this problem can be addressed in the framework of the modified
cosmology or not. The aim is to investigate the the cosmological
Lithium problem  in the background of the modified cosmology, to
check under which conditions and which values for the model
parameter, it can be alleviated.

Our aim in this paper is to explore the implications of Barrow
cosmology on the BBN and derive constraints on Barrow exponent
$\delta$. In this regards, we can check the viability of the
Barrow cosmology by imposing the BBN conditions. The most
important difference between this work and the approach presented
in \cite{John} is that we modify the left-hand side of the
Friedmann equations (the geometry side), while in \cite{John} the
authors modify Friedmann equations in such a way that the Barrow
entropy correction appears as an additional energy density
$\rho_{de}$ and pressure $p_{de}$ which resemble dark energy
components in the right-hand side of the Friedmann equations.
Additionally, in deriving the modified Friedmann equations from
first law of thermodynamics, we neglect the integration constant,
while in \cite{John} the integration constant is interpreted as
the cosmological constant which influences the dynamics of the
universe. Furthermore, we should also note that different
approximations were employed in our approach compared to that in
\cite{John}. We also present the relation between cosmic
time $t$ and temperature $T$ in the early universe and in the
context of the modified Barrow cosmology which is novel and has
not been addressed in the literature. We will see that the
temperature of the early universe increases as the fractal
structure of the boundary of our universe increases.

The structure of the paper is as follows. In the next section, we
apply the first law of thermodynamics, $dE=T_hdS_h+WdV$ on the
apparent horizon, and construct the modified Friedmann equations
based on Barrow entropy. In section \ref{BBN}, we use BBN data in
the background of the modified barrow cosmology to constrain the
Barrow exponent $\delta$. We also present the relation between
time and temperature in the context of Barrow cosmology in this
section. We finish with closing remarks in section \ref{Closing}.
\section{Barrow entropy and Modified Friedman Equations \label{Fried}}
Let us start with a homogeneous and isotropic FRW universe with
line elements
\begin{equation}
ds^2={h}_{\mu \nu}dx^{\mu}
dx^{\nu}+\tilde{r}^2(d\theta^2+\sin^2\theta d\phi^2),
\end{equation}
where $\tilde{r}=a(t)r$, $x^0=t, x^1=r$, and $h_{\mu \nu}$=diag
$(-1, a^2/(1-kr^2))$ represents the two dimensional metric. The
open, flat, and closed universes corresponds to $k = 0, 1, -1$,
respectively. The radius of the apparent horizon, which is a
thermodynamically suitable horizon for the universe, is given by
\cite{Sheyem}
\begin{equation}
    \label{radius}
    \tilde{r}_A=\frac{1}{\sqrt{H^2+k/a^2}}.
\end{equation}
The associated temperature with the apparent horizon is defined as
\cite{MAkbar,Sheyem}
\begin{equation}\label{T}
    T_h=-\frac{1}{2 \pi \tilde
        r_A}\left(1-\frac{\dot {\tilde r}_A}{2H\tilde r_A}\right).
\end{equation}
For $\dot {\tilde r}_A\leq 2H\tilde r_A$, the temperature becomes
negative which is not physically acceptable and hence one should
define $T=|\kappa|/2\pi$. Note that within an infinitesimal
internal of time $dt$  one may assume $\dot{\tilde
    r}_A\ll 2H\tilde r_A$, which physically means that the apparent
horizon radius is fixed. Thus there is no volume change in it and
one may define $T=1/(2\pi \tilde r_A )$ \cite{CaiKim}. The
relation between horizon temperature and Hawking radiation has
been confirmed \cite{cao}.

We further assume the energy content of the Universe is in the
form of a perfect fluid with energy-momentum tensor,
\begin{equation}\label{T1}
T_{\mu\nu}=(\rho+p)u_{\mu}u_{\nu}+pg_{\mu\nu},
\end{equation}
where $\rho$ and $p$ are the energy density and pressure,
respectively and we assume the conservation of energy holds,
namely, $\nabla_{\mu}T^{\mu\nu}=0$. In the background of FRW
universe, this leads to
\begin{equation}\label{Cont}
\dot{\rho}+3H(\rho+p)=0,
\end{equation}
where $H=\dot{a}/a$ is the Hubble parameter. The work density
associated with the volume change, in an expanding universe, is
given by \cite{Hay2}
\begin{equation}\label{Work}
W=-\frac{1}{2} T^{\mu\nu}h_{\mu\nu},
\end{equation}
where in the background of FRW background can be calculated as
\begin{equation}\label{Work2}
W=\frac{1}{2}(\rho-p).
\end{equation}
We also assume the first law of thermodynamics on the apparent
horizon is satisfied and has the form
\begin{equation}\label{FL}
dE = T_h dS_h + WdV,
\end{equation}
where $E=\rho V$ is the total energy of the Universe enclosed by
the apparent horizon, and $T_{h}$ and $S_{h}$ are, respectively,
the temperature and entropy associated with the apparent horizon.
Differentiating $E=\rho V$, yields
\begin{equation}
    \label{dE1}
    dE=4\pi\tilde
    {r}_{A}^{2}\rho d\tilde {r}_{A}+\frac{4\pi}{3}\tilde{r}_{A}^{3}\dot{\rho} dt.
\end{equation}
where $V=\frac{4\pi}{3}\tilde{r}_{A}^{3}$ is the volume enveloped
by a 3-dimensional sphere with the area of apparent horizon
$A=4\pi\tilde{r}_{A}^{2}$. Combining with the conservation
equation (\ref{Cont}), we get
\begin{equation}
    \label{dE2}
    dE=4\pi\tilde
    {r}_{A}^{2}\rho d\tilde {r}_{A}-4\pi H \tilde{r}_{A}^{3}(\rho+p) dt.
\end{equation}
Taking differential form of Barrow entropy (\ref{S}), we find
\begin{eqnarray} \label{dS}
    dS_h&=&
    (2+\delta)\left(\frac{4\pi}{A_{0}}\right)^{1+\delta/2}
    {\tilde
        {r}_{A}}^{1+\delta} \dot{\tilde {r}}_{A} dt.
\end{eqnarray}
Inserting relation (\ref{Work2}), (\ref{dE2}) and (\ref{dS}) in
the first law of thermodynamics (\ref{FL}) and using definition
(\ref{T}) for the temperature, after some calculations, we find
the differential form of the Friedmann equation as
\begin{equation} \label{Fried1}
    \frac{2+\delta}{2\pi A_0 }\left(\frac{4\pi}{A_0}\right)^{\delta/2}
    \frac{d\tilde {r}_{A}}{\tilde {r}_{A}^{3-\delta}}= H(\rho+p) dt.
\end{equation}
Combining with the continuity equation (\ref{Cont}), arrive at
\begin{equation} \label{Fried2}
    -\frac{2+\delta}{2\pi A_0
    }\left(\frac{4\pi}{A_0}\right)^{\delta/2} \frac{d\tilde
        {r}_{A}}{\tilde {r}_{A}^{3-\delta}}=
    \frac{1}{3}d\rho.
\end{equation}
Integration yields
\begin{equation} \label{Frie3}
    \frac{2+\delta}{2-\delta}\left(\frac{4\pi}{A_0}\right)^{\delta/2}
    \frac{1}{2\pi A_0} \frac{1}{\tilde
        {r}_{A}^{2-\delta}}=\frac{\rho}{3},
\end{equation}
where we have set the integration constant equal to zero.
Substituting $\tilde {r}_{A}$ from Eq. (\ref{radius}) we
immediately arrive at
\begin{equation} \label{Fried4}
\frac{2+\delta}{2-\delta}\left(\frac{4\pi}{{A_0}}\right)^{\delta/2}\frac{1}{2\pi
A_0} \left(H^2+\frac{k}{a^2}\right)^{1-\delta/2} = \frac{\rho}{3}.
\end{equation}
The above equation can be further rewritten as
\begin{equation} \label{Fried4}
\left(H^2+\frac{k}{a^2}\right)^{1-\delta/2} = \frac{8\pi G_{\rm
eff}}{3} \rho,
\end{equation}
where the effective Newtonian gravitational constant is defined as
\begin{equation}\label{Geff}
 G_{\rm eff}\equiv \frac{A_0}{4} \left(
 \frac{2-\delta}{2+\delta}\right)\left(\frac{A_0}{4\pi
}\right)^{\delta/2}.
\end{equation}
In this way we obtain the modified Friedmann equation inspired by
Barrow entropy. Thus, starting from the first law of
thermodynamics at the apparent horizon of a FRW universe, and
assuming that the apparent horizon area has a fractal features,
due to the quantum-gravitational effects, we derive the
corresponding modified Friedmann equation of FRW universe with any
spatial curvature. It is important to note that in this case,
there is no dark energy component emerged from the Barrow entropy
that was derived in Ref. \cite{Emm2}. In the limiting case where
$\delta=0$, the area law of entropy is recovered and we have
$A_{0}\rightarrow4G$. In this case, $G_{\rm eff}\rightarrow G$,
and Eq. (\ref{Fried4}) reduces to the standard Friedmann equation
in Einstein gravity.
\section{Constraints on Barrow Exponent inspired by BBN \label{BBN}}
In this section we examine the BBN in the framework of Barrow
cosmology to constrain the Barrow exponent $\delta$.
\subsection{BBN in Barrow cosmology}
The modified Friedmann equation inspired by the Barrow entropy in
a flat universe ($k=0$), can be written as
\begin{eqnarray} \label{firstm}
H=\left ( \frac{8 \pi G_{\rm eff}}{3} \rho \right
)^{\frac{1}{2-\delta}}.
\end{eqnarray}
Substituting $G_{\rm eff} $ from \eqref{Geff} in to above
equation, we get
\begin{eqnarray} \label {hey}
H=\left(\frac{8 \pi \rho}{3}\right)^{\frac{1}{2-\delta}}
\left[\left(\frac{A_{0}}{4}\right)
\left(\frac{2-\delta}{2+\delta}\right)\left(\frac{A_{0}}{4\pi}\right)^{(\delta/2)}\right]^\frac{1}{2-\delta}.
\end{eqnarray}
We can rewrite (\ref {firstm}) in the form
\begin{eqnarray} \label{hz}
    H(T)\equiv Z(T) H_{GR}(T),
\end{eqnarray}
where $ H_{GR}=\sqrt{\frac{8 \pi }{3M_p^2} \rho(T)}$ is the Hubble
function in the standard cosmology and Z(T)is Amplication factor. In the limiting case where
$\delta=0$ we have $A_{0}\rightarrow 4G $, and \eqref {hey} can be rewritten in the form
\begin{eqnarray}
H(T)=\left(\frac{8\pi\rho}{3}\right)^{\frac{1}{2-\delta}}\left[
G\;
\left(\frac{2-\delta}{2+\delta}\right)\left(\frac{G}{\pi}\right)^{\delta/2}\right]^{\frac{1}{2-\delta}},
\end{eqnarray} \
and $Z(T)$ is defined as
\begin{align}\label{amp}
    & Z(T)\equiv \frac{H}{H_{GR}(T)}  \rightarrow
    Z(T) \equiv \notag \\
    & \left[ \left(\frac{8\pi }{3M_p^2} \rho \right)\left(\frac{2-\delta}{2+\delta}\right)\left(\frac{1}{\pi M_p^2}\right)^{\delta/2} \right]^{\frac{1}{2-\delta}}\left(\frac{8\pi }{3M_p^2} \rho\right)^{-1/2} \notag \\
    &=\left[\left(\frac{8\pi }{3M_p^2}\rho\right)^{\delta/2}\left(\frac{2-\delta}{2+\delta}\right)\left(\frac{1}{\pi M_p^2}\right)^{\delta/2} \right]^{\frac{1}{2-\delta}}.
\end{align}
Inserting energy density of relativistic particles which is given
by $ \rho(T)=\frac{\pi^2}{30}g_{*}T^4$ ( $g_{*}\sim10$ is the
effective number of degrees of freedom and $T$ is the temperature)
into Eq. (\ref {amp}) and using $ ( G)^ {-1}=M_{p}^2 $ (in the
units where $\hbar=c=k_B=1$), we find the amplification factor
$Z(T )$ as
\begin{align}\label{zt}
Z(T)\equiv\left(\frac{T}{M_{p}}\right)^{\frac{2\delta}{2-\delta}}
\left[\frac
{2\pi}{3}\sqrt{\frac{g_{*}}{5}}\left(\frac{2-\delta}{2+\delta}\right)^{1/{\delta}}\right]^{\frac{\delta}{2-\delta}}\equiv\eta\left(\frac{T}{M_{P}}\right)^\nu,
\end{align}
where we have defined
\begin{align}\label{eta}
\eta\equiv \left[\frac
{2\pi}{3}\sqrt{\frac{g_{*}}{5}}\left(\frac{2-\delta}{2+\delta}\right)^{1/{\delta}}\right]^{\frac{\delta}{2-\delta}}
\; , \; \; \nu\equiv\frac{2\delta}{2-\delta}.
\end{align}
As expected, for $\delta=0$ (which implies that $\nu=0$), the
amplification factor takes the value $Z(T)=1$, and Einstein
gravity is recovered.

Now we examine the BBN through Barrow cosmology. The primordial $
_{}^{4}\textit{He}$ formation occurs at a time when the
temperature was about $T\sim100 MeV $ and the constituents of the
energy and number density at this time were electrons, positrons,
neutrinos and antineutrinos, plus photons. All these particles are
in thermal equilibrium because of their rapid collisions. Protons
and neutrons were kept in thermal equilibrium through the
interactions
\begin{align}\label{rates}
\nu_{e}+n\leftrightarrow p+e^-, \notag \\
e^++n\leftrightarrow p+\overline{\nu}_{e}, \\
n\leftrightarrow p+\overline{\nu}_{e}+e^- \notag.
\end{align}
The neutron abundance can be estimated through calculating the
conversion rate of neutrons to protons, indicated with
$\lambda_{np}(T)$ and its inverse $\lambda_{pn}(T)$. Hence the
weak interaction rates at enough high temperature is
\begin{equation}\label{tot}
\lambda _{tot}(T)=\lambda _{np}(T)+\lambda _{pn}(T),
\end{equation}
where $\lambda_{np}(T)$ is given by the sum of (\ref {rates})
rates, i.e.,
\begin{eqnarray}\label{1}
\lambda _{np}(T)&=&\lambda _{(n+\nu _{e}\rightarrow
p+e^{-})}+\lambda_{(n+e^{+}\rightarrow p+\bar{\nu
}_{e})}\nonumber\\ &&+\lambda _{(n\rightarrow p+e^{-}+\bar{\nu
}_{e})}.
\end{eqnarray}
While the relation between the rates $\lambda_{np}(T)$ and
$\lambda_{pn}(T)$ is
$\lambda_{np}(T)=e^{{\mathcal{Q}}/T}\lambda_{pn}(T)$, with
$\mathcal{Q}=m_{n}-m_{p}$ the neutron and proton mass difference.
From Eq. (\ref {tot}) one can obtain \cite
{JohnD,Bernstein,Turner}
\begin{align}\label{3}
&   \lambda _{tot}(T)=4AT^{3}(4!T^2+2\times 3!\mathcal{Q}T+2!\mathcal{Q}^2) \notag\\
& \simeq qT^5+ \mathcal{O}\left(\frac{\mathcal{Q}}{T}\right),
\end{align}
with $ A=1.02\times 10^{-3}GeV$ and $q=9.6 \times 10^{-10} GeV^{-4}$.\\
We can estimate the primordial mass fraction of  $ _{}^{4}\textrm{He}$, by using \cite {Turner}
\begin{equation}\label{yp}
Y_{p}\equiv \lambda \frac{2x(t_{f})}{1+x(t_{f})},
\end{equation}
where $\lambda =e^{-(t_{n}-t_{f})/\tau } $, with $t_{f}$ the weak
interactions freeze-out time, $t_{n}$ the nucleosynthesis
freeze-out time, $\tau=\lambda^{-1}_{(n\rightarrow
p+e^{-}+\overline{\nu}_{e})}\simeq887 sec$ the neutron mean
lifetime  \cite {Olive}, and $   x(t_{f})=e^{-\mathcal{Q}/T(t_{f})
}$ is the equilibrium ratio of neutron to proton. The function
$\lambda(t_{f})$ is the fraction of neutrons decaying into protons
inside the range $t\in[t_{f}, t_{n}]$. In any modified cosmology,
the Hubble parameter $H$ will have a deviation from $H_{GR} $, and
hence $T_{f}$ will also deviate from the GR results. Consequently,
the primordial mass fraction $Y_{p}$ will also present a deviation
$\delta Y_{p}$ from the standard cosmological model. From (\ref
{yp}) we have
\begin{align}\label{3}
    lnY_{p}=& ln\lambda+ln2+lnX(t_{f})-ln(1+X(t_{f})) \notag \\
    & \rightarrow \;
    \frac{\delta ln Y_{p}}{\delta T_{f}}=\frac{\delta Y_{p}}{Y_{p} \delta T_{f}}=\frac{\delta}{\delta T_{f}}\left[-\frac{t_{n}-t_{f}}{\tau} \right. \notag \\
    & \left. +ln2-\frac{\mathcal{Q}}{T(t_{f})}-ln(1+X(t_{f})\right] \notag \\
    =&-\frac{2}{\tau T_{f}^{3}}+\frac{\mathcal{Q}}{T_{f}^2}-\frac{\delta X(t_{f})}{\delta T_{f} (1+X(t_{f}))} \notag \\
    &
    =-\frac{2t_{f}}{T_{f}\tau}+\frac{\mathcal{Q}}{T_{f}^2}-\frac{X(T_{f})}{X(T_{f})+1}\,\frac{\mathcal{Q}}{T_{f}^2}.
\end{align}
Thus, we get
\begin{align}\label{yp1}
\frac{\delta Y_{p}}{Y_{p} \delta T_{f}}=& -\frac{2t_{f}}{T_{f}\tau}+\frac{\mathcal{Q}}{T_{f}^2}\left(1-\frac{Y_{p}}{2\lambda}\right) \notag \\
&=\frac{1}{T_{f}}\left[
-\frac{2t_{f}}{\tau}+\frac{\mathcal{Q}}{T_{f}}\left(1-\frac{Y_{p}}{2\lambda}\right)\right].
\end{align}
And hence deviations from the primordial mass fraction $ \delta
Y_{p}$ can be written as
\begin{align}\label{dyp}
\delta
Y_{p}=Y_{p}\left[\left(1-\frac{Y_{p}}{2\lambda}\right)ln\left(\frac{2\lambda}{Y_{p}}-1\right)-\frac{2t_{f}}{\tau}\right]
\frac{\delta T_{f}}{T_{f}}.
\end{align}
Here relations  $t_{f}\simeq T^{-2} $ (because BBN happens at the
radiation dominated epoch),
$ln\left(\frac{1}{X(t_{f})}\right)=ln\left(\frac{2\lambda}{Y_{p}}-1\right)$
and $\frac{X}{X+1}=\frac{Y_{p}}{2\lambda}$ have been used and we
set $ \delta T(t_{n})=0 $ for the reason that $T_{n}$ is fixed by
the deuterium binding energy
\cite{Torres,Lambiase,Lambiasee,Lambiaseee,capoz}.

The observational estimations of the fractional mass $Y_{p}$ of  $
_{}^{4}\textit{He}$ are \cite{Aver}
\begin{align}\label{3}
    Y_{p}=0.2449\pm0.0040 \;,\; \; \; |\delta Y_{p}| < 10^{-4}.
\end{align}
Inserting these into Eq. (\ref{dyp}), the upper bound on ${\delta
T_{f}}/{T_{f}}$ will be extracted as
\begin{align}\label{dtftf}
    \left|\frac{\delta T_{f}}{T_{f}}\right|< 4.7\times10^{-4}.
\end{align}
The freeze-out temperature $T_{f}$ is the temperature at which
particles decouple, and corresponds to the time where $ H=\lambda_
{tot}(T)=qT^5 $, with $H$ is given by (\ref {firstm}), and hence
we have
\begin{align}\label{freeze}
H=& \sqrt{\frac{8 \pi}{3M_{p}^2} \left( \frac{\pi^2g_{*}}{30}T_{f}^4\right)}\; \times \;\eta\left(\frac{T_{f}}{M_{p}}\right)^\nu=qT_{f}^5 \notag \\
& \rightarrow  \; \; \,
T_f^{3-\nu}=\eta\;\frac{2\pi}{3}\sqrt{\frac{\pi
g_{*}}{5}}\frac{1}{qM_{p}^{\nu+1}}.
\end{align}
Following \cite{Anish}, we can examine the freeze-out temperature
in Barrow cosmology. The aforementioned equality allows to compute
the temperature of freeze-out as
\begin{align}\label{Tf}
    T_{f}=M_{p}\left[\eta\;\frac{2\pi}{3}\sqrt{\frac{\pi g_{*}}{5}}\frac{1}{qM_{p}^{4}}\right]^{\frac{1}{3-\nu}}.
\end{align}
Since $ \delta T_{f}=T_{f}-T_{0f}$ , where $T_{0f}\sim0.6 MeV$ \cite {JohnD}, by using equation (\ref {Tf}) we have
\begin{eqnarray}\label{3}
    \left|\frac{\delta T_{f}}{T_{f}}\right|&=&\left|1-\frac{T_{0f}}{T_{f}} \right| \notag\\
    &=&\left| 1-\frac{T_{0f}}{M_{p}}\left[\eta\frac{2\pi}{3}\sqrt{\frac{\pi g_{*}}{5}}\frac{1}{qM_{p}^4}\right]^{-\frac{1}{3-\nu}}
    \right|.
\end{eqnarray}
Inserting $\eta$ from Eq. (\ref {eta}) into the above equation, we
find
\begin{multline} \label{theo}
\left|\frac{\delta T_{f}}{T_{f}}\right|=\left|1-\frac{T_{0f}}{M_{p}}\left[\left(\frac {2\pi}{3}\sqrt{\frac{ g_{*}}{5}}\left(\frac{2-\delta}{2+\delta}\right)^{1/{\delta}}\right)^{\frac{\delta}{2-\delta}}\right.\right.\\
\left. \left.\times\frac{2\pi}{3}\sqrt{\frac{\pi
g_{*}}{5}}\frac{1}{qM_{p}^4}\right]^{-\frac{1}{3-\nu}}\right|,
\end{multline}
where
\begin{align}\label{3}
    3-\nu=\frac{6-5\delta}{2-\delta}.
\end{align}
The constraint on $\delta$ parameter can be found by using Eqs.
(\ref {dtftf}) and (\ref {theo}). In Fig. 1 we plot $ \delta
T_f/T_f$ from (\ref {theo}) vs  $\delta$. The upper bound in
(\ref{dtftf}) is also represented. Therefore, the constraints from
BBN yield
\begin{align}\label{bound}
\delta \simeq 0.01.
\end{align}
This result is consistent with constraint $ 0\leq  \delta\leq 1$,
where the parameter $\delta$ of the Barrow entropy is always
bounded.
\begin{figure}[H]
    \includegraphics[scale=0.8]{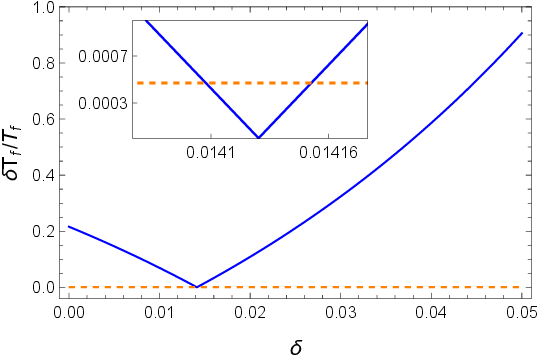}
    \caption{The behavior of $\delta T_f/T_f $ vs $\delta$ in the modified
        Barrow cosmology for early universe. Here $\delta T_f/T_f $ is defined
        in (\ref {theo}), while the observational upper bound of $\delta T_f/T_f $ is given by Eq. (\ref {dtftf}). Constraints from BBN requires $\delta \simeq 0.01$.}
    \label{fig1}
\end{figure}
\subsection{Primordial light elements{ \textit{ $_{}^{4}\textrm{He}$, D} and \textit {Li} } in Barrow cosmology}
In this section we derive the constraints on Barrow exponent
$\delta$ in a different approach in which we probe the effects of
Barrow cosmology on the primordial abundances of light elements,
i.e., Deuterium $_{}^{2}\textit{H}$, Tritium $_{}^{7}\textit{Li}$,
and Helium $_{}^{4}\textit{He}$. We use the observational data for
the abundances of aforementioned light elements to derive the
bound on Barrow parameter. The main idea is to replace the
amplication factor $Z(T)$ instead of the usual one entering the
primordial  light elements abundances, which is related to the
neutrinos species effective number \cite {Luciano}. We observe
that in the standard cosmological model one simply has $Z=1$.
Modified gravity or the presence of extra light particles such as
neutrinos, may deviate Z from unity, in which case the modified
Z-factor takes the form \cite{Anish,Luciano,Boran}
\begin{align}
    Z_{\nu}=\left [ 1+\frac{7}{43} (N_{\nu}-3)\right ]^{1/2},
\end{align}
where $N_{\nu} $ is the number of neutrinos species. The
baryon-antibaryon asymmetry $\eta_ {10}$, plays an important part
in this analysis \cite{Adv,Simha}. Since we focus on the effects
of Barrow cosmology in BBN, we set $N_{\nu}=3 $, rejecting the
possibility that in our framework deviations of $Z$ from unity
appears due to the presence of extra particle degrees of freedom.

In what follows, we follow the approach of \cite{Anish,Sahoo}. We
remind the main features:

-\textit {$_{}^{4}\textrm{He}$ abundance}- The first step of
production of $_{}^{4}\textit{He}$ consists in generating
$_{}^{2}\textit{H}$ from a neutron and a proton. After that
$\textit {D}$ is converted into Tritium and $_{}^{3}\textit{He}$.
The sequence of nuclear reactions that produce Helium
$_{}^{4}\textit{He}$ are
\begin{align}
    n + p &\rightarrow \, ^2\text{H} + \gamma \label{eq:32},\\
    ^2\text{H} + ^2\text{H} &\rightarrow \, ^3\text{He} + n, \label{eq:33} \\
    ^2\text{H} + ^2\text{H} &\rightarrow \, ^3\text{H} + p.
\end{align}
The Helium $_{}^{4}\textit{He}$ is finally produced through the following processes
\begin{align}
    ^2\text{H} + ^3\text{H} &\rightarrow \, ^4\text{He} + n,\\
    ^2\text{H} + ^3\text{He} &\rightarrow \, ^4\text{He} + p. \label{eq:35}
\end{align}
Numerical best fit constrains the primordial abundance of
$_{}^{4}\textit{He}$ to be \cite {Kneller,Annu}
\begin{equation} \label{bestfit}
    Y_p = 0.2485 \pm 0.0006 + 0.0016 \left[ (\eta_{10} - 6) + 100 (Z - 1) \right],
\end{equation}
where in our case $Z$ is given by Eq. (\ref {zt}). Here, the
baryon density parameter $\eta_ {10}$ is defined as
\cite{Adv,Simha}
\begin{equation} \label {et}
    \eta_{10} \equiv 10^{10}\eta_B \equiv 10^{10} \frac{n_B}  {n_{\gamma}} \simeq
    6,
\end{equation}
where $\eta_B \equiv{n_B} /{n_{\gamma}}$ is the baryon to photon
ratio \cite {Wamp}. For $Z=1$, we recover the standard result
predicted by BBN for the $_{}^{4}\textit{He}$ abundance
$(Y_p)|_{GR}=0.2485\pm 0.0006$. The observational data on
$_{}^{4}\textit{He}$ abundance and setting $\eta_ {10}= 6$ implies
that $Y_p=0.2449\pm 0.004$ \cite{Brain}. Consistency between
aforementioned value of $Y_p$ and Eq.(\ref{bestfit}) gives
\begin{equation}
    0.2449 \pm 0.0040 = 0.2485 \pm 0.0006 + 0.0016 \left[ 100(Z - 1)
    \right].
\end{equation}
Thus we can fix $Z$ as
\begin{equation} \label {zhe}
    Z = 1.0475 \pm 0.105.
\end{equation}
- \textit {$_{}^{2}\textit{H}$ abundance}- Deuterium is produced
from the process $ n + p \rightarrow \, ^2\text{H} + \gamma $ .
Following the same approach as above, Deuterium abundance can be
determined from the numerical best fit in \cite {Adv}, giving
\begin{equation} \label {zde}
    y_{Dp} = 2.6(1 \pm 0.06) \left(\frac{6} {\eta_{10} - 6(Z - 1)}
    \right)^{1.6}.
\end{equation}
As before the values $Z=1$ and $\eta_ {10}=6 $ yield the results
in standard cosmology $ Y_{D_p}|_{GR}=2.6\pm 0.16 $. Equating the
observational constraint on \textit{D} abundance  $ Y_{D_p}=2.55\pm 0.03$
\cite {Brain}, with Eq.\eqref {zde}, one gets
\begin{equation}
    2.55 \pm 0.03 = 2.6(1 \pm 0.06) \left(\frac{6} {\eta_{10} - 6(Z - 1)}
    \right)^{1.6}.
\end{equation}
And hence the constraints on $Z$ is
\begin{equation} \label {zobs}
    Z = 1.062 \pm 0.444,
\end{equation}
which is partially overlaps with the constraint from $_{}^{4}\textit{He}$ abundance in Eq. (\ref{zhe}). \\
- \textit{$_{}^{7}\textit{Li}$ abundance}- The $\eta_ {10}$
parameter (\ref {et}) on the one hand successfully fits the
$_{}^{4}\textit{He}$, $D$ and other light elements abundances, and
on the other hand is somehow inconsistent with observed
$_{}^{7}\textit{Li}$ abundance. In fact, the ratio of predicted
value of $_{}^{7}\textit{Li}$ abundance in the standard
cosmological model to the observed one lies in the range
\cite{Boran}
\begin{align*}
    \frac{\text{Li}|_{GR}}
    {\text{Li}|_{obs}}
    \in [2.4 - 4.3].
\end{align*}
Quite unexpectedly, neither standard BBN nor any other model are
able to fit this low abundance of $_{}^{7}\textit{Li}$. This
puzzle is known as \textit {Lithium problem} \cite {Boran}. By
demanding consistency between observational bounds on Lithium
abundance $Y_{Li}=1.6\pm0.3$ \cite {Brain}, and the numerical best
fit for $_{}^{7}\textit{Li}$ abundance
\begin{equation}
    y_{Li} = 4.82 (1 \pm 0.1)\left[\frac{\eta_{10} - 3(Z - 1)} {6} \right]^{2},
\end{equation}
where $Z$ can be fixed as
\begin{equation} \label {lit}
    Z = 1.960025 \pm 0.076675.
\end{equation}
This constraint does not overlap with the results for
$_{}^{2}\textit{H}$ abundance in Eq.(\ref {zobs}), and
$_{}^{4}\textit{He}$ abundance in Eq.(\ref {zhe}).
\subsection{Discussion on the results}
Let us discuss the obtained results, which we derived for the
bound on the Barrow parameter $\delta$. In Fig. 2 we plot (\ref
{zt}) taking into account the bounds presented in (\ref {zhe}).
As we can see, the parameter $\delta$ lies in the range \\
\begin{equation} \label {dhe}
0\lesssim \delta \lesssim 0.001.
\end{equation}
\begin{figure} [H]
    \includegraphics[scale=0.88]{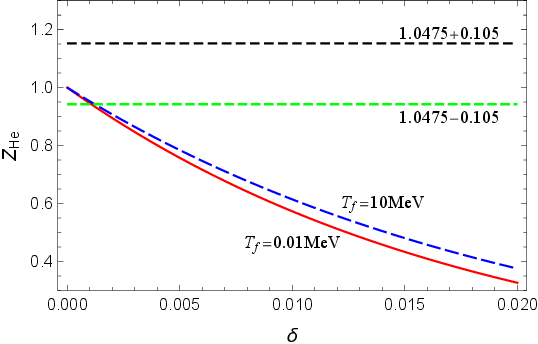}
    \caption{$Z_{_{}^{4}\textrm{He}}$ vs $\delta$.
The experimental range is reported in (\ref {zhe}). We have fixed
$\eta_{10}=6$ and varied the temperature of freeze-out in the
range $T_f=[0.01,10] MeV$.}
    \label{fig2}
\end{figure}
We can observe the results for Deuterium (using Eq.(\ref {zobs}))
in Fig. 3. It is seen that the corresponding range for $\delta$ is
 \begin{equation} \label {deut}
 0\lesssim \delta \lesssim 0.008.
 \end{equation}
As we can see, the range of $\delta$ for Deuterium (\ref {deut})
overlaps with the range for $_{}^{4}\textit{He}$ given by Eq.
(\ref {dhe}).
 \begin{figure} [H]
 \includegraphics[scale=0.88]{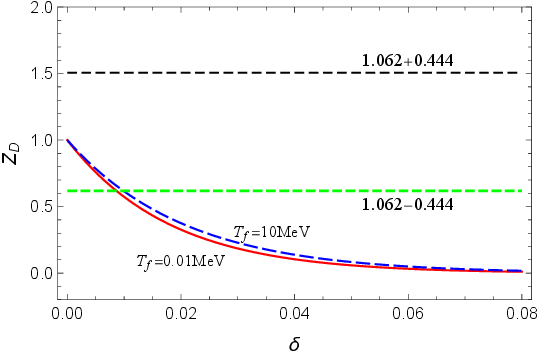}
 \caption{$Z_{_{}^{2}\textrm{H}}$ vs $\delta$.
The experimental range is reported in (\ref {zobs}). We have fixed
$\eta_{10}=6$ and varied the temperature of freeze-out in the
range $T_f=[0.01,10] MeV$.}
 \label{fig3}
 \end{figure}
In the case of Lithium, Fig. 4 shows that the range of $\delta $
for $_{}^{7}\textit{Li}$ does not overlap with the constraints on
$\delta$ that $_{}^{4}\textit{He}$ and $_{}^{2}\textit{H}$
abundances provided in Eqs. (\ref {dhe}) and (\ref {deut}).
 \begin{figure}[H]
 \includegraphics[scale=0.88]{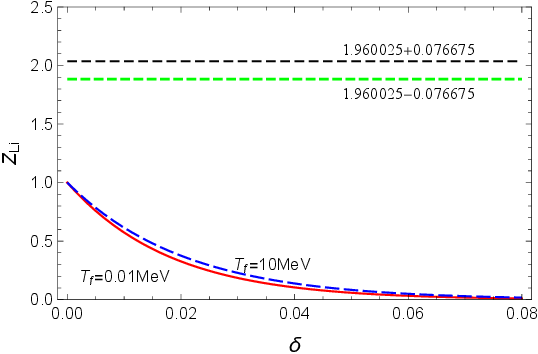}
 \caption{$Z_{Li}$ vs $\delta$.
The experimental range is reported in (\ref {lit}). We have fixed
$\eta_{10}=6$ and varied the temperature of freeze-out in the
range $T_f=[0.01,10] MeV$.}
 \label{fig4}
\end{figure}
\section{Relation between cosmic time and temperature}
The condition of thermal equilibrium in early universe results to
entropy conservation in a co-moving volume \cite {Weinberg}
   \begin{align}\label{3}
        s(T)a^3=\rm constant,
    \end{align}
where $s(T)$ is the entropy per unit co-moving volume.  Taking the
derivative of the above equation with respect to time gives
    \begin{align}
        \dot{s}(T)a^3+3\dot{a}a^2s(T)=0
        \mapsto \frac{ds(T)}{dt}a^3=-3\dot{a}a^2s(T).
    \end{align}
Using modified Friedmann equation (\ref{firstm}), we arrive at
    \begin{align}\label{3}
        \frac{ds(T)}{dt}=-3\left(\frac{8\pi G_{\rm eff}}{3}
        \rho\right)^{1/2-\delta}s(T).
    \end{align}
We can write the above equation in the form
    \begin{align}\label{3}
        dt=-\frac{ds(T)}{3s(T)}\left(\frac{8\pi G_{\rm eff}}{3}
        \rho\right)^{1/\delta-2}.
    \end{align}
    And hence the cosmic time $t$ can be written as
    \begin{align}\label{zxcv}
        t=-\frac{1}{3}\int_{}^{} \frac{{s}'(T)}{s(T)}\left(\frac{8\pi G_{\rm eff}}{3} \rho\right)^{1/\delta-2}
        dT,
    \end{align}
    where the prime stands for the derivative with respect to temperature $T$.\\
     In particular, during any era in which the dominant constituent of the universe is a
     highly relativistic ideal gas, the entropy and energy densities are given by \cite {Weinberg}
      \begin{align}\label{3}
     s(T)=\frac{2\mathcal{N}a_B T^3}{3}\;
     \end{align}
     \begin{align}\label{3}
      \rho(T)=\frac{\mathcal{N}a_B T^4}{2},
     \end{align}
      (where $\mathcal{N} $ is the number of particles and antiparticles and each spin state separately) \cite {Weinberg}.
      Inserting $\frac{{s}'(T)}{s(T)}=\frac{3}{T} $ and aforementioned definition for $\rho$ into the Eq. \eqref {zxcv}, we obtain
    \begin{align}\label{3}
        t=-\left(\frac{8\pi G_{\rm eff}}{3}\right)^{1/\delta-2}\int_{}^{}\frac{1}{T}\left(\frac{\mathcal{N}a_B T^4}{2}\right)^
        {1/\delta-2}dT.
    \end{align}
We can rewrite the above equation in the following form
    \begin{align}\label{3}
        t=-\left(\frac{8\pi G_{\rm eff}\mathcal{N}a_{B}}{6}\right)^{1/\delta-2}\int_{}^{}T^
        {\frac{6-\delta}{\delta-2}}dT.
    \end{align}
    Integrating yields
    \begin{align}\label{tTbarrow}
    t=\frac{1}{T^{4/(2-\delta)}}\left(\frac{8\pi G_{\rm eff}\mathcal{N}a_{B}}{6}\right)^{1/\delta-2}\frac{2-\delta}{4}+ \rm constant.
   \end{align}
   The above equation is the relation between the cosmic time $t$ and temperature $T$ in  the context of the modified Barrow cosmology.
   For $\delta=0$ we find $t=\frac{1}{T^2}\sqrt{\frac{3}{16\pi G\mathcal{N}a_{B}}} $, which is the result of standard cosmology \cite {Weinberg}.

   Let us note that we  have used the modified Friedmann equation which has been derived in the units where $\hbar=k_B=c=1 $.
   On the other hand, the right hand side of Eq. \eqref{tTbarrow} should have the dimension $[\,t\,]$.
   If we take into account all the constants in deriving the modified Friedmann equations, the effective gravitational constant
   in Eq. \eqref{Geff} takes the form
   \begin{equation}
    G_{\mathrm{eff }} \equiv \frac{A_{0}}{4}\left(\frac{2-\delta}{2+\delta}\right)
    \left(\frac{A_{0}}{4 \pi}\right)^{\delta / 2}{c^{2-\delta}\over \alpha_B \Gamma_B},
    \end{equation}
    where we have defined
    \begin{equation}
        \alpha_B\equiv k_B\left( c^3\over \hbar \right)^{1+(\delta/2)}\quad \,,\,\,\quad \Gamma_B\equiv {\hbar c\over
        k_B}.
    \end{equation}
Thus, one can easily check that the right side of Eq.
\eqref{tTbarrow} has dimension $[\,t\,]$. In early universe, when
the constituents of the universe were photons, plus three species
of neutrinos and three of antineutrinos, plus electrons and
positrons, we have $\mathcal {N} =43/4$, so that Eq.
\eqref{tTbarrow} yields, in cgs units,
 \begin{equation}
    t=\left( {T\over10^{10}K} \right)^{4\over
    \delta-2}F(\delta)+\rm {constant}.
 \end{equation}
 Where we have defined
 \begin{eqnarray}
F(\delta)&=&\left[\left( \frac{2-\delta}{2+\delta}  \right)\times (9.2 \times 10^{-88})^{\delta/2}\right]^{1\over \delta-2}\nonumber\\
&&\times {2-\delta \over 4} (0.278)^{1/(\delta-2)}
 \end{eqnarray}
 Again for $\delta=0$ we have $F(\delta)\to 0.994\; sec $ and the
 result of standard cosmology is restored  \cite {Weinberg}
\begin{eqnarray}
t\to \;0.994 \;sec\;\left[ {T\over 10^{10}K} \right]^{-2}.
 \end{eqnarray}
In Fig. 5 we plot the behavior of the temperature $T$ as a
function of cosmic time $t$ for different values of Barrow
parameter $\delta$. We observe that as the fractal structure
increases, the temperature of the early universe increases too.
\begin{figure}[H]
\includegraphics[scale=0.88]{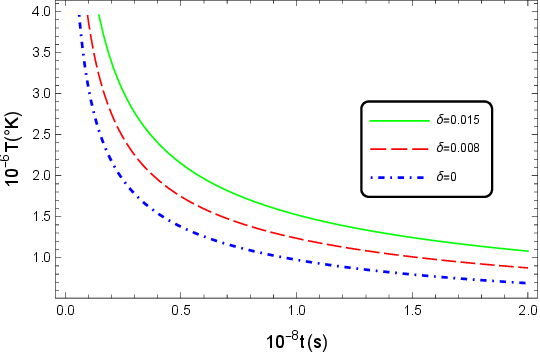}
 \caption{The behavior of $T $ vs $t$ in modified
 Barrow cosmology for early universe.}
 \label{fig5}
 \end{figure}
\section{Closing remarks \label{Closing}}
In this paper we have used BBN data in order to find the bounds on
the free parameter $\delta$ of Barrow entropy, which is an
extended entropy relation originating from the incorporation of
quantum gravitational effects on the structure of a black hole,
parameterized by $\delta$. Barrow entropy leads to modified
Firedmann equations containing extra terms. These extra terms
should be sufficiently small in order not to spoil BBN epoch. We
can summarize our results in this paper as follows:

(i) We first presented an analysis of the BBN epoch in the Barrow
cosmology and calculated the deviations of the temperature of
freeze-out in comparison to the result of standard cosmology. We
have used observational upper bound on $\left | \frac{\delta
T_f}{T_f} \right |$ which is the allowed deviation from GR, in
order to extract the bound on Barrow exponent $\delta$. We have
shown that the BBN constraint imposes a value for the Barrow
exponent $\delta$ of the order $\delta \simeq\mathcal{O}(0.01)$. As
expected this result shows that the deformation from
Bekenstin-Hawking expression should be small.

(ii) We have also explored the consequences of Barrow cosmology on
the primordial light elements formation. The analysis of the light
elements $_{}^{4}\textit{He}$, $_{}^{2}\textit{H}$,$
_{}^{7}\textit{Li}$ shows that the range of  variability of
$\delta$, using the bounds (\ref {zhe}), (\ref {zobs}) and (\ref
{lit}), does not overlap.

(iii) We have presented the relation between cosmic time $t$ and
the temperature $T$ of the universe in the modified Barrow
cosmology. We have then plotted the behavior of temperature as a
function of cosmic time $t$ for different $\delta$. The result
shows that as the fractal structure increases, the temperature of
the early universe increases, too.

Following the present study, we can use BBN data in order to
impose constraints on the free parameter of other entropy
corrected cosmological models such as modified cosmology inspired
by zero-point length, or the modified Kaniadakis cosmology. We can
also investigate the relation between time and temperature in
these models. We leave these for the future investigations.
\acknowledgments{We thank Shiraz university Research Council. The
work of A. Sheykhi is based upon research funded by Iran National
Science Foundation (INSF) under project No. 4022705.}


\begin{thebibliography}{99}
\bibitem {Bardeen} J. M. Bardeen, B. Carter, S. Hawking, \textit{The four laws of black hole mechanics}, Commun. Math. Phys. \textbf{31}, 161170 (1973).
\bibitem {Bekenstein}J. D. Bekenstein, \textit {Black holes and entropy}, Phys. Rev. D \textbf {7}, 2333 (1973).
\bibitem {Hawking} S. Hawking, \textit{Black Holes and Thermodynamics }, Phys. Rev. D \textbf {13}, 191 (1976).
\bibitem {Jacobson} T. Jacobson, \textit{Thermodynamics of spacetime: the Einstein equation of state}, Phys. Rev. Lett. \textbf {75}, 1260 (1995).
\bibitem {Verlinde} E. Verlinde, \textit{On the origin of gravity and the laws of Newton} JHEP \textbf {1104}, 029 (2011).
\bibitem {Padman} T. Padmanabhan, \textit{Gravity and the thermodynamics of horizons}, Phys. Rep \textbf {406}, 49 (2005).
\bibitem {Padmann} T. Padmanabhan,\textit{Thermodynamical aspects of gravity: new insights}, Rept. Prog. Phys. \textbf {73}, 046901 (2010).
\bibitem {Eling}  C. Eling, R. Guedens, and T. Jacobson, \textit {Nonequilibrium thermodynamics of spacetime}, Phys. Rev. Lett. \textbf {96}, 121301 (2006).
\bibitem {Akbar} M. Akbar and R. G. Cai, \textit{Friedmann equations of FRW universe in scalar-tensor gravity, $f(R)$ gravity and first law of thermodynamics}, Phys. Lett. B \textbf {635}, 1 (2006).
\bibitem {MAkbar} M. Akbar,  R. G. Cai, \textit{Thermodynamic behavior of the Friedmann equation at the apparent horizon of the FRW universe},  Phys. Rev. D \textbf{75}, 084003 (2007).
\bibitem {Cai}  R. G. Cai and L. M. Cao, \textit{Unified first law and the thermodynamics of the apparent horizon in the FRW universe}, Phys. Rev. D \textbf {75}, 064008 (2007).
\bibitem {Asheykhi} A. Sheykhi, B. Wang, R. G. Cai, \textit{Thermodynamical properties of apparent horizon in warped DGP braneworld}, Nucl. Phys. B \textit {779}, 1 (2007).
\bibitem {Sheykh} A. Sheykhi, B. Wang and R. G. Cai, \textit{Deep connection between thermodynamics and gravity in Gauss-Bonnet braneworlds}, Phys. Rev. D textbf {76}, 023515 (2007).
\bibitem {Paranj} A. Paranjape, S. Sarkar and T. Padmanabhan, \textit{Thermodynamic route to field equations in Lanczos-Lovelock gravity}, Phys. Rev. D \textbf {74}, 104015 (2006).
\bibitem {Jamil} M. Jamil, E. N. Saridakis and M. R. Setare, Phys. Rev. \textit{Thermodynamics of dark energy interacting with dark matter and radiation}, D \textbf {81}, 023007 (2010).
\bibitem {Caiii}  R. G. Cai and N. Ohta, \textit {Horizon thermodynamics and gravitational field equations in Horava-Lifshitz gravity}, Phys. Rev. D textbf {81}, 084061 (2010).
\bibitem {Wang} M. Wang, J. Jing, C. Ding and S. Chen, \textit{First law of thermodynamics in IR modified Horava-Lifshitz gravity}, Phys. Rev. D \textbf {81}, 083006 (2010).
\bibitem {Setare} M. Jamil, E. N. Saridakis and M. R. Setare, \textit{The generalized second law of thermodynamics in Horava-Lifshitz cosmology}, JCAP \textbf {1011}, 032 (2010).
\bibitem {Fan}  Z. Y. Fan and H. Lu, \textit {Thermodynamical first laws of black holes in quadratically-extended gravities}, Phys. Rev. D \textbf{91}, 064009 (2015).
\bibitem{Tsheykhi} A. Sheykhi,\textit{ Modified Friedmann equations from Tsallis entropy},  Phys. Lett. B \textbf {785}, 118 (2018).
\bibitem{JohnD} John D. Barrow, \textit{The area of a rough black hole}, Phys. Lett. B \textbf{808}, 135643 (2020).
\bibitem{John} J. D. Barrow, S. Basilakos , E. N. Saridakis, \textit {Big-Bang nucleosynthesis constraints on Barrow entropy}, Phys. Lett. B \textbf {815}, 136134 (2021).
\bibitem{HawkingS} S. W. Hawking, \textit{Spacetime foam}, Nucl. Phys. B \textbf {144}, 349 (1978).
\bibitem{Hooft} G. Hooft, \textit{On the quantum structure of a black hole}, Nucl. Phys. B \textbf {256}, 727 (1985).
\bibitem{Shey1}  A. Sheykhi, \textit{Barrow entropy corrections to Friedmann equations}, Phys. Rev. D \textbf{103}, 123503 (2021).
\bibitem{Shey2}  A. Sheykhi, \textit{Modified cosmology through Barrow entropy}, Phys. Rev. D \textbf{107}, 023505 (2023).
\bibitem {Luciano}  G.G. Luciano, \textit{Primordial big bang nucleosynthesis and generalized uncertainty principle} Eur. Phys. J. C \textbf{81}, 1086 (2021).
\bibitem{Anish} A. Ghoshal, G. Lambisae, \textit{Constraints on Tsallis cosmology from big bang nucleosynthesis and dark matter freeze-out}, [arXiv:2104.11296].
\bibitem {Boran}  S. Boran and E. O. Kahya,\textit{Testing a dilaton gravity model using nucleosynthesis}, Adv. High Energy Phys. \textbf {1}, 282675 (2014).
\bibitem{Sheyem}  A. Sheykhi, \textit{Friedmann equations from emergence of cosmic space}, Phys. Rev. D {\bf87}, 061501(R) (2013).
\bibitem{CaiKim} R. G. Cai and S. P. Kim,\textit{First Law of Thermodynamics and Friedmann Equations of Friedmann-Robertson-Walker Universe}, JHEP {\bf0502}, 050 (2005).
\bibitem{cao}  R. G. Cai, L. M. Cao, Y. P. Hu,\textit{Hawking Radiation of Apparent Horizon in a FRW Universe}, Class. Quant. Grav. {\bf26} 155018 (2009).
\bibitem{Hay2} S. A. Hayward, \textit{Unified first law of black-hole dynamics and relativistic thermodynamics}, Class. Quant. Grav. {\bf 15}, 3147 (1998).
\bibitem{Emm2} E. N. Saridakis, \textit{Modified cosmology through spacetime thermodynamics and Barrow horizon entropy,} JCAP {\bf07}, 031 (2020).
\bibitem{Bernstein}J. Bernstein, L. S. Brown, G. Feinberg, \textit{Cosmological helium production simplified }, Rev. Mod. Phys. \textbf {61} (1989).
\bibitem{Turner} E.W. Kolb, M.S. Turner, \textit{The Early Universe, Addison Wesley Publishing Company}, (1989).
\bibitem{Olive}  K. A. Olive, et, al., (Particle Data groups), Ch. Phys. C \textbf {38}, 0900001 (2014).
\bibitem{Torres} D. F. Torres, H. Vucetish, A. Plastino, \textit {Early universe test of nonextensive statistics}, Phys. Rev. Lett. \textbf{79}, 1588 (1997).
\bibitem{Lambiase} G. Lambiase, \textit{Lorentz invariance breakdown and constraints from big-bang nucleosynthesis}, Phys. Rev. D \textbf {72}, 087702 (2005).
\bibitem{Lambiasee} G. Lambiase, \textit{Constraints on massive gravity theory from big bang nucleosynthesis}, J. Cosmol. Astropart. Phys. \textbf {10}, 028 (2012).
\bibitem{Lambiaseee} G. Lambiase, \textit{Dark matter relic abundance and big bang nucleosynthesis in Horava's gravity}, Phys. Rev. D \textbf {83}, 107501 (2011).
\bibitem{capoz} S. Capozziello, G. Lambiase, E. Saridakis, \textit{Constraining $f(T)$ teleparallel gravity by big bang nucleosynthesis: $f (T)$ cosmology and BBN}, Eur. Phys. J. C \textbf {77}, 576 (2017).
\bibitem{Aver} E. Aver, K.A. Olive, and E.D. Skillmann, \textit {The effects of $He$ I $\lambda10830$ on helium abundance determinations}, JCAP \textbf {07}, 011 (2015).
\bibitem {Adv} G. Steigman, \textit{Neutrinos and big bang nucleosynthesis}, Adv. High Energy Phys. \textbf {1}, 268321 (2012).
\bibitem {Simha} V. Simha, G. Steigman,\textit {Constraining the early-Universe baryon density and expansion rate}, JCAP \textbf {06}, 016 (2008).
\bibitem {Sahoo} S. Bhattacharjee, P.K. Sahoo, \textit{Big bang nucleosynthesis and entropy evolution in $f (R, T)$ gravity}, Eur. Phys. J. Plus \textbf{135}, 350 (2020).
\bibitem {Kneller} J. P. Kneller, G. Steigman, \textit{BBN for pedestrians}, New J. Phys. \textbf{6}, 117 (2004).
\bibitem {Annu} G. Steigman, \textit{Primordial nucleosynthesis in the precision cosmology era}, Annu. Rev. Nucl. Part. Sci. \textbf{57}, 463 (2007).
\bibitem {Wamp} N. Jarosik, et al. \textit{Seven-year wilkinson microwave anisotropy probe (WMAP) observations: Sky maps, systematic errors, and basic results}, Astrophys. J. Suppl. \textbf{192}, 18 (2011).
\bibitem {Brain} B. D. Fields, K. A. Olive, T.H. Yeh, C. Yung, \textit {Big-bang nucleosynthesis after Planck}, JCAP \textbf{03}, 010 (2020).
\bibitem{Weinberg} S. Weinberg, \textit {Cosmology, New York: Oxford University press}, (2008).
\end{thebibliography}
\end{document}